\documentclass[aps,pra,reprint,superscriptaddress,floatfix,showpacs]{revtex4-1}
\usepackage{xcolor}
\usepackage{subfigure}
\usepackage{cancel}

\usepackage{float}
\usepackage{amsmath}
\usepackage{amssymb}
\usepackage{empheq} 
\usepackage{braket}

\usepackage{bm}
\usepackage{graphicx}
\usepackage{hyperref}
\usepackage{natbib}
\usepackage{siunitx}
\usepackage{soul} 

\hypersetup{
    colorlinks,
    linkcolor={blue!50!blue},
    citecolor={blue!50!blue},
    urlcolor={blue!
    80!black}
}

\begin{document}

\title[A low cost plasmonic platform for photon emission engineering of two dimensional semiconductors]{A low cost plasmonic platform for photon emission engineering of two dimensional semiconductors} 

\author{Anuj Kumar Singh}
\author{Kishor K Mandal}
\author{Yashika Gupta}
\author{Abhay Anand VS}
\affiliation{Laboratory of Optics of Quantum Materials, Department of Physics, IIT Bombay, Mumbai - 400076, India}
\author{Lekshmi Eswaramoorthy}
\author{Brijesh Kumar}
\affiliation{Laboratory of Optics of Quantum Materials, Department of Physics, IIT Bombay, Mumbai - 400076, India}
\author{Abhinav Kala}
\affiliation{Department of Condensed Matter Physics and Material Science, Tata Institute of Fundamental Research, Homi Bhabha Road, Mumbai, 400005 India}
\author{Saurabh Dixit}
\affiliation{Laboratory of Optics of Quantum Materials, Department of Physics, IIT Bombay, Mumbai - 400076, India}
\author{Venu Gopal Achanta}
\email{achanta@tifr.res.in}
\affiliation{Department of Condensed Matter Physics and Material Science, Tata Institute of Fundamental Research, Homi Bhabha Road, Mumbai, 400005 India}
\author{Anshuman Kumar}
\email{anshuman.kumar@iitb.ac.in}
\affiliation{Laboratory of Optics of Quantum Materials, Department of Physics, IIT Bombay, Mumbai - 400076, India}
\date{\today}
\keywords{plasmonics, WSe$_2$, PTFE, nanocone, photo luminescence}

\begin{abstract}
Although the field of 2D materials has democratized materials science by making high quality samples accessible cheaply, due to the atomically thin nature of these systems, an integration with nanostructures is almost always required to obtain a significant optical response. Traditionally, these nanostructures are fabricated via electron beam lithography or focused ion beam milling, which are expensive and large area fabrication can be further time consuming. In order to overcome this problem, we report the integration of 2D semiconductors on a cost-effective and large area fabricated nanocone platform. We show that the plasmon modes of our nanocone structures lead to  photoluminescence enhancement of monolayer WSe$_2$ by about eight to ten times compared to the non-plasmonic case, consistent with finite-difference time-domain simulations. Excitation power-dependent measurements reveal that our nanocone platform enables a versatile route to engineering the relative exciton trion contributions to the emission. 
\end{abstract}


\maketitle

\section{Introduction}
Monolayers of transition metal dichalcogenides (TMDCs) are optically active direct bandgap semiconductors~\cite{10.1038/natrevmats.2017.33} which have been shown to be promising candidates for optoelectronic applications~\cite{Mak2016,Pospischil2016} such as sensing~\cite{Lee2018}, photovoltaics~\cite{Wang2018}, and quantum information~\cite{Liu2019}. The integration of these 2D semiconductor TMDCs with nanostructures can not only strengthen the light-matter interaction~\cite{Schneider2018,Krasnok2018} but also help engineer their optical response for various applications~\cite{Yan2020,bib1,Li2016,You2020}. Although the field of 2D materials has democratized materials science~\cite{CastellanosGomez2016} by making high quality samples accessible cheaply, due to the atomically thin nature of these systems, an integration with nanostructures is almost always required to obtain a significant optical response~\cite{Krasnok2018}. Traditionally, these nanostructures are fabricated via electron beam lithography or focused ion beam milling, which are expensive and large area fabrication can be further time consuming~\cite{Nianqiang}.  In order to overcome this problem, we report the integration of 2D semiconductors on a cost-effective and large area fabricated nanocone platform. We fabricate Polytetrafluoroethylene (PTFE) nanocone structures and decorate it with a gold (Au) film, enabling it to behave like a plasmonic antenna array. We show that the plasmon modes of our nanocone structures lead to  photoluminescence (PL) enhancement of monolayer WSe$_2$ by about eight to ten times compared to the non-plasmonic case. PL enhancement is further verified via Finite-difference time-domain (FDTD) simulations. Excitation power-dependent as well as the time-dependent measurements reveal that our nanocone platform enables a versatile route to engineering the relative exciton trion contributions to the emission.
\section{Methods}
\begin{figure*}
    \includegraphics[width=0.8\textwidth]{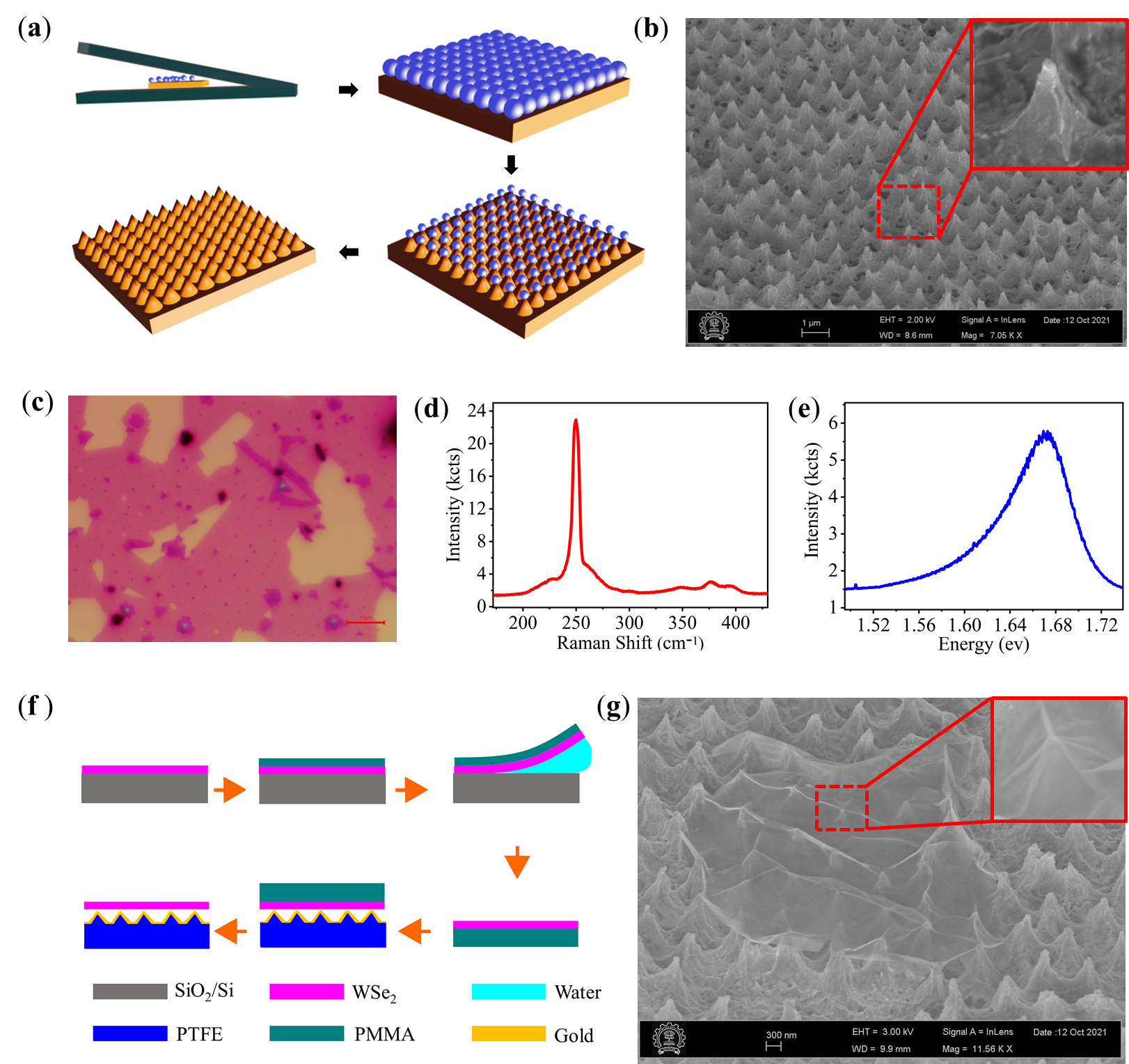}
    \caption{\textbf{Fabrication of the TMDC integrated plasmonic nanocone structures. }{{\textbf{(a)}} Process Flow diagram for the fabrication of PTFE cone array    {\textbf{(b)}} SEM image of large scale PTFE  nanocone array after etching (inset showing zoomed view of the cone) {\textbf{(c)}} Optical microscopy image of the large area CVD grown WSe$_2$ monolayer {\textbf{(d)}} Raman spectrum confirming the monolayer {\textbf{(e)}} PL spectra of the monolayer WSe$_2$ {\textbf{(f)}} Process flow diagram for the wet transfer method {\textbf{(g)}} SEM image of the WSe$_2$ monolayer transferred over the cone array (inset showing the WSe$_2$ over single cone)}}
    \label{fab}
\end{figure*}

\begin{figure*}
    \includegraphics[width=1.03\textwidth]{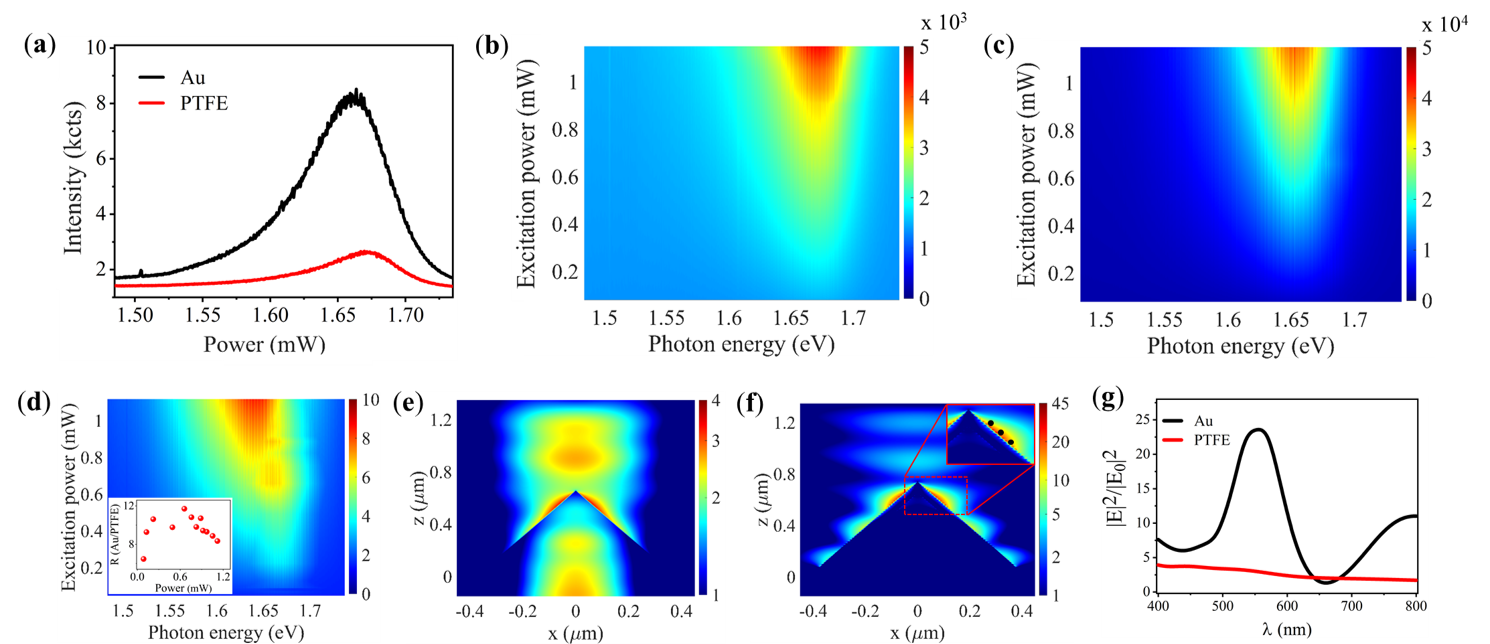}
    \caption{\textbf{PL enhancement at various excitation powers and the explanation with FDTD simulation. }{ {\textbf{(a)}} micro-PL spectra of the transferred WSe$_2$ over PTFE and Au nanocone structures. PL intensity with the variation of the excitation energy {\textbf{(b)}} For the PTFE cones {\textbf{(c)}} Au cones and {\textbf{(d)}} ratio (R) between Au to the PTFE {(inset shows the line plot for ratio R for maximum of the PL peak intensity (Au/PTFE) with the excitation power)}. FDTD simulation for the calculation of Field enhancement {\textbf{(e)}} For PTFE cone {\textbf{(f)}} For the Au cone {\textbf{(g)}} Line spectra for the electric field enhancement showing resonance.  }}
    \label{pl-color}
\end{figure*}

\begin{figure*}
    \includegraphics[width=0.8\textwidth]{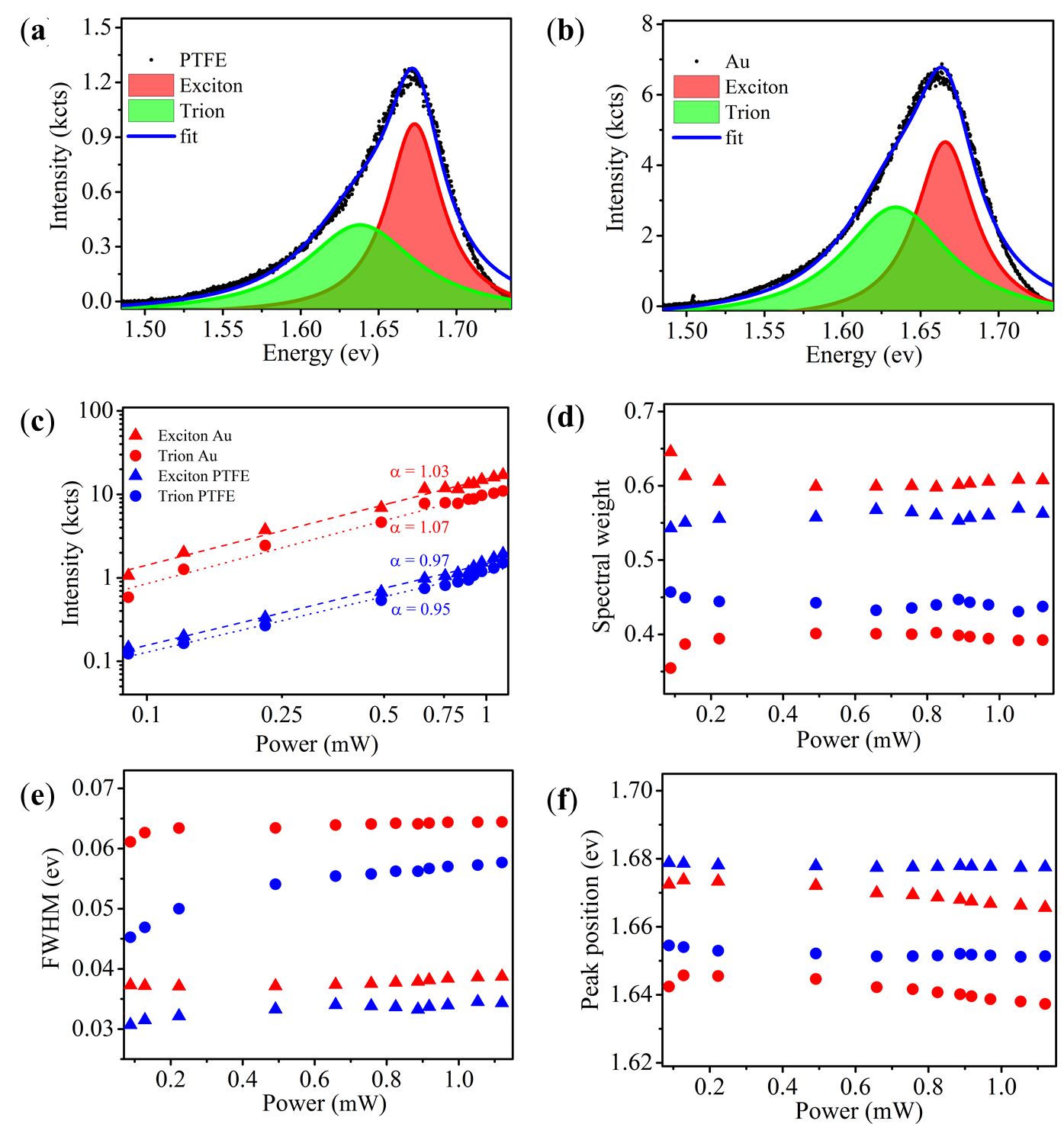}
    \caption{\textbf{ Exciton-trion dynamics with excitation power after fitting with double lorentzian. }{ PL spectra at the 532 nm pump laser  fitted with double lorentzian  for {\textbf{(a)}}  WSe$_2$ transfered over PTFE cone {\textbf{(b)}} WSe$_2$ transfered over Au cone. {\textbf{(c)}} PL intensity at various  excitation energy with linear fitting, $\alpha$ denotes the slope {\textbf{(d)}} Intensity spectral weight for the exciton and trion  {\textbf{(e)}} FWHM of the exciton and trion  {\textbf{(f)}} Peak position for  the exciton and trion, in PTFE and Au cones   }}
    \label{et-dynamics}
\end{figure*}
\emph{Fabrication of nanocone array--}
Large-scale fabrication of nanocones was carried out via a colloidal lithography approach~\cite{Zhang2009,Singh2021}. Firstly, PTFE substrates, purchased from sigma aldrich (GF46787259-1EA) were cleaned using ethanol for five minutes in an ultrasonicator and then washed with de-ionized water. Next, the substrates were treated with an oxygen plasma for two minutes to make them hydrophilic. Polystyrene (PS) microbeads were deposited on this modified PTFE surface using convective self-assembly~\cite{Sun2010,Prevo2004}. The process flow diagram for the fabrication of the cone array is shown in Figure~\ref{fab}(a), where two glass slides were aligned to form $1^\circ$ wedge angle for the deposition of the PS microparticles and a diluted $1\%$ suspension of PS micro-particle was then injected into the wedge. This assembly was carefully kept undisturbed for two hours, allowing the suspension liquid to evaporate, leaving behind PS microbead monolayer covering the substrate. The surface of the monolayer of PS microparticles was then exposed to an oxygen plasma to reduce the size of the deposited PS microparticles and in turn etch out the exposed PTFE film below to form nanocones with the etching time. This etching was carried out via reactive ion etching in an oxygen plasma maintained at 0.2~mbar pressure and 40~sccm gas flow rate with 100~W RF power. The etching parameters were optimized to obtain an array of PTFE nanocones surmounted by residues of PS microparticles. These residues were removed by rinsing the sample in ethanol.  Figure~\ref{fab}(b) shows the post etching scanning electron microscopy (SEM) image of the as-fabricated PTFE cone array on PTFE substrate, with the inset showing a high-resolution image of a single nanocone.  A 70~nm thick Au layer was then sputtered over these cone arrays using an AJA international, Inc/ Orion Sputter PHASE. The sputtering parameters used were 50~W power, 7.4x$10^{-7}$ torr chamber vacuum, and a deposition rate of 6~nm/min. A chemical vapor deposition (CVD) grown WSe$_2$ monolayer was transferred over the as-fabricated cone array using polymethyl methacrylate (PMMA) assisted wet transfer method~\cite{13}.

\emph{Monolayer TMDC growth and integration--}We optimized CVD process to obtain large area growth of monolayer of around cm-scale. The optical microscopy image of the as-grown WSe$_2$ monolayers is shown in Figure~\ref{fab}(c). Raman spectra with characteristic ${\rm E_{2g}^{1}}$ peak 249.82~$cm^{-1}$ with the absence of ${\rm B_{2g}}$ peak at 303~$cm^{-1}$ and PL spectra with emission peak at 1.67~eV in Figure~\ref{fab}(d,e) confirm the presence of monoloyer~\cite{Feng_2018,Liu2015}. This CVD-grown WSe$_2$ monolayer was then transferred over SiO$_2$ substrate using the PMMA assisted wet transfer method as demonstrated in Figure~\ref{fab}(f) over both the PTFE cone and Au coated PTFE cones. Wet transfer method relies on weakening the interaction of 2D material with substrate by using solutions like water~\cite{Zhang_2017} or NaOH/KOH~\cite{13} while the strong interaction  2D material with the top coated polymer (PMMA in our case) remains unaffected by these solutions. The SEM image in Figure~\ref{fab}(g) shows the as-transferred monolayer over the cone array. These monolayers form a tent-like structure due to the tip of the cone as  shown in the inset of Figure~\ref{fab}(g).

\emph{FDTD Simulation--}
To simulate the excitonic emission in WSe$_2$ monolayer integrated with Au and PTFE nanocone array, FDTD numerical method (using Ansys/Lumerical commercial simulation software) was employed.  In the simulation setup, the system is enclosed inside  a 1000 x 1000 x 4500 nm FDTD boundary. Mesh size of the whole simulation was fixed to 2~nm. PL enhancement consists of two factors, excitation enhancement and emission enhancement factor. Excitation enhancement is calculated as the ratio of electric field intensity in presence and absence of the plasmonic nanocone given by $ \rm F = {|E|^2}/{|E_{o}|^2}$, where $\rm|E|$ and $\rm|E_o|$ are the electric field magnitudes with and without the plasmonic structure. For emission enhancement, the exciton in the monolayer was simulated as a horizontal dipole source, sitting 5 nm on the top of the cone surface. Emission enhancement  is  calculated using Eq.~\ref{q}~\cite{Bharadwaj2007}:
\begin{equation}
   \rm Q = \frac{P_{rad}}{(P_{tot}+(1-\eta)P_{rad}^{o}/\eta)}\label{q}
\end{equation}
where, $\rm {P_{rad}}$ is the far field radiated power which is collected over numerical aperture (NA) of the microscope (NA= 0.4) in presence of the nanocone structures. $\rm {P_{tot}}$ and $\rm {P_{rad}^{o}}$ are the total power radiated by the dipole with and without the structure, respectively and ${\rm \eta}$ corresponds to the intrinsic PL quantum yield of the WSe$_2$, taken with typical range of values from literature ~\cite{Eggleston2018, Kim2019,JaverzacGaly2018} as 0.001, 0.01 , 0.015, and 0.03. 

For our system, we draw a comparison between PL enhancement factors obtained for only PTFE and Au covered PTFE cones to explain the experimental results.

\emph{Optical characterisation--}
Raman spectroscopy was used to confirm the presence of  WSe$_2$ monolyers. Raman spectra for the as-fabricated samples were recorded using HR800-UV confocal micro-Raman spectrometer with the help of 100x objective with an excitation source of 532~nm laser light having 1.1~mW power. Raman spectrometer was initially calibrated with standard Raman peak of crystalline silicon at 520.7~cm$\textsuperscript{-1}$. Acquisition time for Raman scattered light collection was 20 seconds. PL measurements were carried out by using a custom made PL setup using 532~nm excitation source, collected with a 20X Mitutoyo Plan Apo NIR Infinity Corrected Objective. Spectra were recorded using a Kymera 328i Andor spectrometer.   

\section{Results and discussion}
 We measured the emission from monolayer TMDC integrated with both PTFE and Au coated PTFE nanocone arrays to understand the difference between TMDC interaction with plasmonic and dielectric antennae platforms. The standard PL spectrum of the WSe$_2$ coated PTFE and Au cones recorded with a 532~nm laser excitation in Figure~\ref{pl-color}(a) shows about eight to ten times enhancement in the PL intensity with Au coating. We attribute this enhancement in PL emission of monolayer WSe$_2$ in case of Au coated cones to the plasmonic properties of Au. We found that this enhancement is strongly dependent on the excitation power, hence we performed systematic PL measurements with varying laser fluences for the two cases. Figure~\ref{pl-color} (b, c) clearly shows higher PL intensity for Au coated cones at all excitation powers. The color map of the ratio of PL intensity of the two cases (Au coated/PTFE) is presented in Figure~\ref{pl-color} (d), which shows a eight to ten times enhancement for low fluence to high fluence, respectively for Au coated cones as compared to the dielectric PTFE ones {(inset shows the line plot where the PL enhancement is approximately 8--10 times)} .

\emph{Plasmonic enhancement--} If the plasmon resonance of the plasmonic-TMDC system matches with the excitation frequency, the excitation rate of the TMDC will be enhanced. Further, a plasmon resonance at the particular PL frequency can enhance the emission rate~\cite{Anonymous1946,Wang2016,Salehzadeh2015}. This means that the PL enhancement in the plasmonic- TMDC system contains two terms, that are the excitation and emission enhancement~\cite{Bharadwaj2007}. We calculated both excitation and emission enhancement~\cite{Bharadwaj2007} using  Lumerical FDTD simulation (see Methods section) for both types of nanocones. In Figure~\ref{pl-color}(e) and (f), the excitation enhancements for the PTFE and Au cones are shown. A clear enhancement of more than ten times for positions close to the cone-tip can be observed for the Au cone as compared to the dielectric one. We further calculated emission enhancement for a dipole placed in close proximity of the tip, that is at $x = 70$, $80$ and $90$~nm, { denoted by black dots in the inset of the Figure~\ref{pl-color}{(f)}}, as discussed in the Methods section. {For the further confirmation of the plasmonic resonance at the excitation wavelength and emission wavelength the absolute square of the electric field at $\rm x = 70 $ nm is plotted for both PTFE and Au cones in Figure~\ref{pl-color}{(g)}. In PTFE cone case there is no resonance seen while in Au case there is about 20 times electric field intensity enhancement at the excitation wavelength and around 10 times at emission wavelength.} If we account for the internal quantum yield, these simulations show that for positions close to the cone tip, an enhancement of $\approx 3$ to $\approx 6.5$ times can be observed for different values of quantum yield in  Au coated cone as compared to the PTFE cone (see Table~\ref{tabel:PL}, which is in the same range as the experimentally measured PL enhancement. 
 
\begin{table}
\caption{Theoretically calculated PL enhancement ratios (Au/PTFE) at positions, $x = 70, 80$ and $90$~nm  for different values of $\eta$. 
\label{tabel:PL}}
\begin{tabular}{|c|c|c|c|c|}
\hline
\multicolumn{1}{|l|}{x (nm)} &\multicolumn{1}{l|}{${\eta}$ = 0.001} & \multicolumn{1}{l|}{${\eta}$ = 0.01} & \multicolumn{1}{l|}{${\eta}$ = 0.015} & \multicolumn{1}{l|}{${\eta}$ = 0.03}\\ \hline
70                         &4.96  & 5.06  & 3.84      &     3.1                         \\ \hline
80                       & 5.82    & 4.5   &   4        &   3.01                      \\ \hline
90                      & 6.57     & 4.19 &   4.49      &   3.3                         \\ \hline
\end{tabular}
\end{table}

\emph{Excitation dependent PL enhancement--} To explain the nonlinear PL enhancement as discussed above, we fitted the PL spectra for the PTFE and Au cases with a double Lorentzian as shown in Figure~\ref{et-dynamics}(a) and (b) respectively. This provides information about the relative exciton and trion contributions and their role in the PL enhancement. We plotted the integrated PL intensity for the exciton and trion as a function of excitation fluence for both the samples and observed a linear relation as shown in Figure~\ref{et-dynamics}(c). The slopes for the Au cones for exciton and trion intensities are found to be 1.03 and 1.07, respectively, while for PTFE cones, they were 0.97 and 0.95, respectively. This suggests that we are in a linear regime where excitation power and the exciton and trion intensities are proportional. At higher excitation powers, we might encounter exciton-exciton interactions which can result in a sublinear dependence~\cite{doi:10.1021/acs.nanolett.9b02431}. We further plotted the intensity spectral weight $\rm I_m/(I_E+I_T)$ for excitons and trions in Figure~\ref{et-dynamics}(d), where subscripts `$\rm E$' represent exciton, `$\rm T$' represents trion and `$\rm m$' stands for the E or T. This quantity provides information about the electrostatic charge neutrality of the monolayer.  We can see that the spectral weight for the excitons in Au and PTFE is approximately  60\% and 55\% respectively, while the trion spectral weight in Au and PTFE is approximately 40\% and 45\% respectively. {It is known that the trion recombination pathway is mostly nonradiative~\cite{Lien2019}.} From here, it is clear that the population of excitons is increasing while trions are decreasing in the case of the Au. This increased population of the excitons in Au coated cones can also contibute to the PL enhancement as compared to the bare PTFE cones~\cite{Tao2018}. {The PL enhancement has previously been reported due to substrate dependent changes ~\cite{Buscema2014,Kwon2018} and hot electron doping \cite{Wen2022}. In our case, from the calculated local fields shown in Figs. 2(e) and 2(f), there is an order of magnitude enhancement in cones deposited with Au compared to without Au coating.}

Next we measured the features of the exciton and trion peaks, full width at half maxima (FWHM) and peak positions with increasing excitation powers. As shown in Figure~\ref{et-dynamics}(e), the FWHM for both these cases increases slightly with increase in excitation power. This can be attributed to an increased dephasing of these excitons and trions due to interaction with the increasing background of free carriers. Further, the increased FWHM can also occur due to heating of the samples due to laser irradiation, which causes an increased phonon excitation and the exciton-phonon coupling results in broadening of the spectra~\cite{Lundt2018,Kaupmees2020}. This photo induced heating effect is more dominant for the plasmonic case versus non plasmonic, as reported elsewhere~\cite{Najmaei2014}.  Next, we observe a small redshift in the exciton and trion peak positions with increasing excitation power.  Usually, the bandgap renormalization effect and carrier screening induced exciton binding energy lowering change oppositely with increased excitation power~\cite{Park2017,Cunningham2017}. However, since we observe only the redshift of the peak, the former effect dominates in our case. However, in our range of powers, these shifts are small compared to the respective peak positions, hence should not impact the performance of nanophotonic devices built using our platform.

\begin{figure*}
    \includegraphics[width=\textwidth]{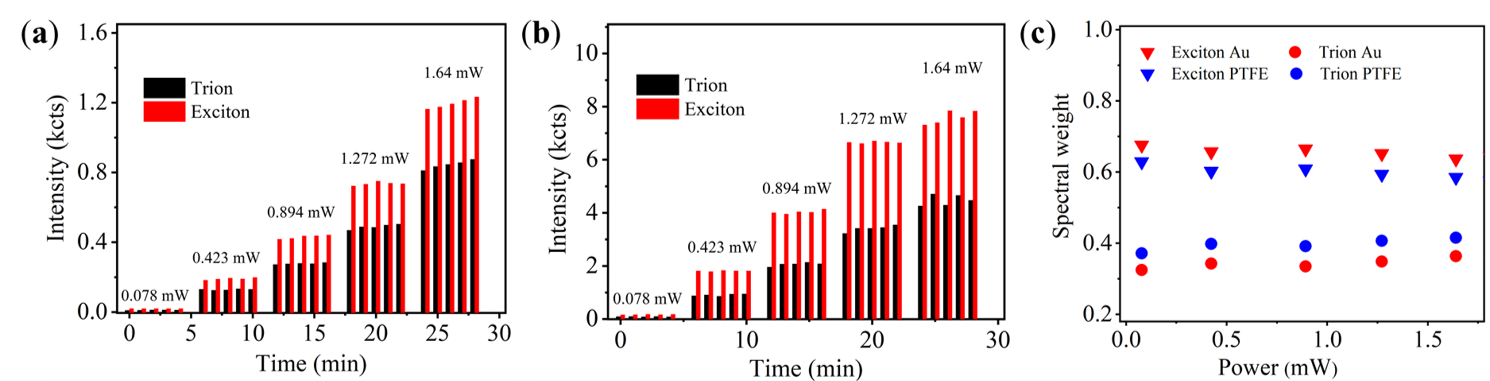}
    \caption{\textbf{ Time dependent analysis for the exciton-trion dynamics.}{ Intensity of the integrated peak as a function of time for {\textbf{(a)}} PTFE  and {\textbf{(b)}} Au. {\textbf{(c)} Shows the variation of intensity spectral weight with excitation power.}  }}
    \label{time}
\end{figure*}

Next we measured the time evolution of the excitonic and trionic features by recording PL for several minutes at fixed values of pump power. We plot the intensity of the integrated PL peak as a function of time and the excitation power in Figure~\ref{time}. For both PTFE and Au cones, the intensity did not change much with the time (see Figure~\ref{time}(a and b)). We also plotted the intensity spectral weight for the fourth minute and observed that the excitonic spectral weight has the same trend as that was observed previously in Figure~\ref{et-dynamics}(d), hence confirming the stability of our system with time.\\

\section{Conclusion}
In summary, we presented the integration of 2D TMDCs with a low-cost and large area plasmonic nanocone array platform. We studied the exciton dynamics and resulting PL enhancement when the nanocones are covered with a plasmonic material. We observed a PL enhancement of eight to ten times which was explained using FDTD simulations and higher excitonic spectral weight. Other than the traditional avenues of plasmon enhanced optoelectronics~\cite{Yan2020}, our platform would enable unique applications to deterministic strain and dielectric screening based periodic modulation of the optical response of 2D semiconductors~\cite{Peng2020}, including the development of quantum emitter arrays~\cite{Chakraborty2019,Yu2021}, exciton funneling based devices~\cite{Feng2012,Moon2020,Lee2020,Su2022} and the observation of dark excitons~\cite{Chand2022,Rahaman2021}.
\\
\emph{Acknowledgement--} A.K.S acknowledges financial support from Industrial Research and Consultancy Center (IRCC) IIT Bombay. A.K. acknowledges funding support from the Department of Science
and Technology via the grants: SB/S2/RJN-110/2017,
ECR/2018/001485 and DST/NM/NS-2018/49. We acknowledge the Centre of Excellence in Nanoelectronics (CEN) at IIT Bombay for providing fabrication facilities.

\bibliography{ref.bib}

\end{document}